\documentclass[prl,twocolumn,aps]{revtex4}
\usepackage{graphicx}
\usepackage{epsfig}
\begin{document}

\draft

\title{Quantum Communication Through a Spin-Ring with Twisted Boundary Conditions}

\author{S. Bose}
\affiliation{Department of Physics and Astronomy, University
College London, Gower St., London WC1E 6BT, UK}

\author{B.-Q. Jin}
\author{V.\ E.\ Korepin}
\affiliation{C.N.\ Yang Institute for Theoretical Physics, State
  University of New York at Stony Brook, Stony Brook, NY 11794-3840}
\date{\today}

\begin{abstract}
We investigate quantum communication between the sites of a
spin-ring with twisted boundary conditions. Such boundary conditions can be
 achieved by a magnetic flux through the ring.
We find that a
non-zero twist can improve communication through finite odd
numbered rings and enable high fidelity multi-party quantum
communication through spin rings (working near perfectly for
rings of 5 and 7 spins). We show that in certain cases, the twist
results in the complete blockage of quantum information flow to a
certain site of the ring. This effect can be exploited to
interface and entangle a flux qubit and a spin qubit without
embedding the latter in a magnetic field.
\end{abstract}

%\pacs{Pacs No: 03.67.-a, 03.65.Ud, 32.80.Lg}

\maketitle

%\begin{multicols}{2}
Recently, quantum communication through unmodulated spin chains
has developed into a topic of much interest
\cite{Bose02,Subrahmanyam04,Christandl04,Yung03,Osborne04,Amico04,Burgarth04,Clark04,Eisert04,Plenio04}.
Here the word ``spin chains" means chains of qubits whose interactions are always on. The communication is achieved by simply placing a quantum state at
one end of the chain and waiting for an optimized time to let this
state propagate to the other end with a high fidelity. In such a
scheme, the chain acts as a non-photonic quantum channel for short
distance communications between, say, two quantum computers. It
helps to avoid interfacing solid state, trapped atomic and other
quantum computation hardware with optics because quantum channels
can now be made by the same hardware. This is highly desirable for
short distance quantum communications \cite{wineland02}, as
interfacing back and forth between electron/atoms and photons may
be complex, time consuming and not worth when the distance is
really short. A positive feature of such schemes is the
non-necessity of switching interactions on and off between
individual pairs of spins of the chain in a time synchronized
manner. This helps to prevent the accumulation of errors with
increasing number of switchings. For a small number of spins
connecting nearby parties, there are even more advantages. Then it
suffices to keep only the finite number of sites in which the
spins are located free from decoherence (isolated from the
environment). This may be much easier than keeping the decoherence
low for the entire length of a channel through which an
information bearing electron or atom passes. The results for quantum communication using spin chains also apply
to qubit chains where all interactions can be switched on and off simultaneously, but it is difficult to switch them on or off individually. In those cases, the
results for spin chains typically help in investigating the best quantum communication that can be achieved by switching the group of interactions on and off only once.

\begin{figure}
\begin{center}
\includegraphics[width=3in, clip]{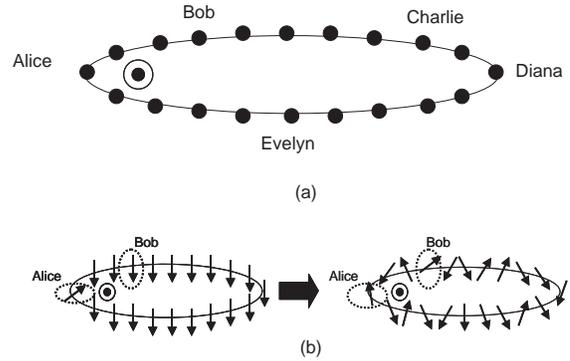}
 \caption{Part (a) shows the scenario for multi-party quantum communication using a spin ring. Alice, Bob, Charlie, Diana and Evelyn are five parties
 which have access to distinct spins in the ring. Any pair of parties would like to communicate with high fidelity. Part (b) shows a way in which the above can be achieved. If Alice wants Bob to be able to receive 
 a state sent by her with high fidelity,
 she puts a certain flux through the spin ring, as shown by concentric circles. By making another choice of flux, Alice can make Diana the candidate to receive a state with high fidelity.}
\label{flux1}
\end{center}
\end{figure}

      So far
\cite{Bose02,Subrahmanyam04,Christandl04,Yung03,Osborne04,Amico04,Burgarth04,Clark04,Eisert04}
(with the exception of Refs.\cite{Osborne04,Burgarth04,Plenio04}), the emphasis has been
to communicate a quantum state between two parties
 through a spin-chain and achieve the highest possible fidelity over the highest possible distance. However, it
would be even more interesting if a single spin chain could be
used as a bus to connect a larger number parties (i.e., a larger
number of quantum computers) such as Alice, Bob, Charlie, Diana
and Evelyn, as shown in Fig.\ref{flux1}(a), and high fidelity
quantum communication could be performed between {\em any pair} of
such parties when required. In this paper, we show that such a
task can be accomplished near-perfectly for 5 and 7 parties using
a closed spin-chain (a ring) with twisted boundary conditions.
This boundary condition is achieved by a magnetic flux through the ring
\cite{yang,bohm},
we illustrated this  in Fig.\ref{flux1}(b). Even though the separation between the
communicating parties is small in this case, such a configuration
is still very useful. Spins belonging to each party could be
interacting with both of its nearest neighbors, yet by tuning the
flux appropriately, one can select any pair of parties along the
ring to perform near perfect quantum communication, this is a new feature
which was absent for open boundary conditions. To achieve the
same otherwise, one would have to have $^5C_2$ or $^7C_2$
separately switchable interactions for the cases of $5$ and $7$
parties respectively. Here not only are there far fewer (namely
$5$ or $7$) interactions, they are also allowed to be
simultaneously on. We also point out the advantages gained for
higher odd $N$ rings by the application of a flux. For certain
even $N$ rings, we show that Alice can also use the twist as a
switch to controllably block the flow of quantum information to a
specially situated Bob (this does not, however, hold for a
generally situated Bob). We show how this interesting effect can
be used to interface and entangle flux and spin qubits without
directly embedding the latter in the magnetic field of the former.
The coupling between the spin and the flux qubit is easily
switched on and off by switching the interactions between the
spins in the ring on and off.

   As a model for the ring, we shall
use spin $\frac{1}{2}$ Heisenberg ferromagnet (i.e. XXX) chain.
The Hamiltonian of the model can be written as:
\begin{equation}
H_{XXX}=-J \sum_{i=1}^N
({\vec \sigma}_i \cdot
{\vec \sigma}_{i+1})-B\sum_{i=1}^N \sigma_i^z.\label{xxxh}
\end{equation}
We can make a Jordan-Wigner transformation and represent the
model as spinless interacting fermions, see for example \cite{takahashi}.
This reformulation of the model naturally leads to twisted boundary
conditions, see \cite{yang}.
% Indeed let us consider a spinless
%fermion carrying a charge $-q$ on a ring of length $N$. Let us
%imagine
% a magnetic flux $\hbar c \Phi /q$ threads
% the ring. Now a spinless fermion going around the ring will pick up a
% phase $\Phi$.
 This is the famous
Aharonov-Bohm (A-B) effect, see \cite{bohm}. 
%The exact magnitude of $q$ remains undetermined by the Jordan-Wigner 
%transformation and will have to be empirically
%determined for the specific system of spins under consideration from the
% observed magnitude of the A-B phase. 
The A-B phase leads to the replacement of periodic
boundary conditions by twisted boundary conditions \cite{ss,yang}
\begin{equation}
|\Psi(j+N)\rangle =e^{i\Phi} |\Psi(j)\rangle.
\end{equation}
Here $j$ is a space coordinate.
So we shall consider twisted boundary conditions. They can
improve quantum communication. Below it will be convenient for us to change the notation $\Phi =2\pi f$
and use a scaled flux $f$.

  The general scheme for quantum communications is as follows.
  The ring is a ferromagnetic Heisenberg
  XXX spin chain with nearest neighbor interactions and is
  initialized in its ground state as in \cite{Bose02}. Alice first
  chooses the party (among Bob, Charlie, Diana and Evelyn) to which she wants to
  communicate. For example, in Fig.\ref{flux1}(b), she has chosen Bob. Then she applies
an appropriate flux $\hbar c \Phi_{\scriptsize{\mbox{BOB}}} /q$ through the
ring. After this, she places the quantum state to be transmitted
at her site. After waiting for a pre-calculated optimal time, Bob
retrieves the state from the chain with a high fidelity. In
Ref.\cite{Bose02}, it has been shown that the fidelity of quantum
communications through the channel under consideration is an
increasing function of the absolute value of the transition
amplitude of $f^N_{r,s}$ of an excitation from the site $r$ to $s$
of an $N$ spin chain. Moreover, it has also been shown that Alice
could transmit entanglement of magnitude $|f^N_{r,s}|$ from the
$r$-th to the $s$-th site of the chain by placing half of a
maximally entangled state on the $r$-th site. In this paper,
thus, we will primarily be interested in examining the
optimization of $|f^N_{r,s}|$ by application of a twist/flux
$\hbar c \Phi /q$.

The eigenstates of the ring with twisted boundary conditions are
\begin{equation}
|\tilde{m}\rangle=\frac{1}{\sqrt{N}}\sum_{j=1}^N e^{i
\frac{2\pi}{N}(m+f) j}|{\bf j}\rangle, \label{eigenv}
\end{equation}
where $f=\Phi/2\pi$ is a fraction between $0$ and $1$ which
quantifies the twist, and $m=1,2,...N$.
The energies of the states $|\tilde{m}\rangle$ are given by
$E_m=-4 J\cos\{{\frac{2\pi}{N}(m+f)}\}-J(N-4)-B(N-2)$.
Correspondingly, the quantum transition amplitude from the $s$-th
site to the $r$-th site at a time $t$ is given by
\begin{eqnarray}
f^N_{r,s}(t) &=& \sum_{m=1}^N \langle {\bf r}|\tilde{m}\rangle
\langle \tilde{m}|{\bf s}\rangle e^{-iE_m t} \nonumber\\
&=& e^{i\frac{2\pi}{N}(r-s)f} {\cal F}(r-s, u_m) \label{dct}
\end{eqnarray}
where, $u_m=\exp(-iE_m t)$ and ${\cal F}(r-s,
u_m)=(1/N)\sum_{m=1}^{N} u_m \exp\{i(2\pi/N)(r-s) m\}$ is the
$(r-s)$-th element of the inverse discrete Fourier transform of
the vector $\{u_m\}$. With the help of identity
\begin{eqnarray}
\exp(iz\cos \theta)=\sum_{n=-\infty}^{\infty} i^n J_n(z) e^{-i n
\theta},
\end{eqnarray}
Eq.~(\ref{dct}) can further be written in the form
\begin{eqnarray}
f^N_{r,s}(t)&=&  e^{-i(4J+2B)t} i^{d}\sum_{k=-\infty}^{\infty}
J_{d-kN}(\beta) i^{-kN} e^{i 2\pi f k} \nonumber\\
&= &  e^{-i(4J+2B)t} \Bigl( i^{d} \sum_{k=0}^{\infty} e^{i 2\pi
({N\over 4}-f)
k} J_{d+kN}(\beta) \nonumber\\
&& + i^{d'} e^{i 2\pi f} \sum_{k=0}^{\infty} e^{i 2\pi ({N\over
4}+f) k } J_{d'+kN}(\beta)\Bigr), \label{fn}
\end{eqnarray}
where $\beta=4Jt$, $d=r-s$ and $d'=N-d$. We will only study the
quantity $\xi=|f^N_{r,s}(t)|$, as it is the entanglement which can
be shared between two sites by transmitting half of an entangled
state through the spin ring channel \cite{Bose02} (the fidelity,
which is an increasing function of $\xi$, has similar behavior).
From Eq.~(\ref{fn}), one immediately obtains that for $N\to
\infty$, $\xi=|J_{r-s}(\beta)|$ and results will be exactly the
same as that of an untwisted ring \cite{Bose02}. However, the
situation is different for $N$ finite and Eq.~(\ref{fn}) can be
used to estimate the optimal twist and time for obtaining the
maximal fidelity of communication between any two sites. We now
proceed to numerically investigate the various consequences of
Eq.~(\ref{fn}) for different $N$.

We first examine very simple odd $N$ rings. Without any twist
Ref.\cite{Bose02}), they have a significantly lower $\xi$ than
even $N$ rings of similar length. This is because, for odd $N$
rings, there are always a pair of equidistant reception sites
$r_1$ and $r_2$ from any $s$-th site from which Alice intends to
transmit her state. Therefore, the amplitude of the excitation and
the distributable entanglement is equally divided among these
sites. This is changed by a twist, which gives a net-momentum
(proportional to twist parameter $f$) to the spin-wave in one
direction. This can also be seen from following evolution equation
of state:
\begin{equation}
|\tilde{m} (t)\rangle=\frac{1}{\sqrt{N}}\sum_{j=1}^N e^{i
\frac{2\pi}{N}(m+f) j -i E_m t}|{\bf j}\rangle, \label{ev}
\end{equation}
and related equation with the replacement $j\to N-j$ and $m\to
N-m$ in Eq.~(\ref{ev}). More specifically, the pair of states
(state $m=p$ and state $m=N-p$ with $p$ taking any value of
$1,\cdots, N-1$) will have the same transmission velocity but
different direction for $f=0$. However, this relation will be
broken for non-zero $f$. Then nearly maximal entanglement
($\xi\sim 1$) can be be distributed between any pair of sites of
small odd $N$ rings such as $N=5$ and $N=7$. These are tabulated
in Table \ref{table1}. From the table, it is clear that up to two
decimal places, maximal entanglement can be transmitted from any
site to any other site of pentagonal and septagonal chains.

\begin{table}
\begin{tabular}{|c|c|c|c|c|c|c|}
  \hline
  & \multicolumn{3}{c}{N=5}\vline &\multicolumn{3}{c}{N=7} \vline\\
  \hline
  $d$ & $f$ & $\beta$ & $\xi$ & $f$ & $\beta$ & $\xi$\\
  \hline
  1 & -0.25 & 1214.3 & 0.9998 & -0.25 & 4365.0 & 0.9997\\
  2 & -0.25 & 162.51 & 0.9999 & 0.25 & 1942.6 & 0.9994\\
  3 & 0.25 & 162.51 & 0.9999 & 0.25 & 3500.4 & 0.9996\\
  4 & 0.25 & 1214.3 & 0.9998 & -0.25 & 3500.4 & 0.9996\\
  5 & n/a & n/a & n/a &-0.25 & 1942.6 & 0.9994 \\
  6 & n/a & n/a & n/a & 0.25 & 4365.0 & 0.9997 \\
\hline
  \end{tabular}
\caption{Table showing the maximum entanglement $\xi$
distributable between pairs of sites of $N=5$ and $N=7$ spin
rings. The various variables are the distance $d=r-s$ between the
sending and the receiving sites,
  the twist parameter $f$ and the scaled time $\beta$ at which the given values of $\xi$ are attained.}
  \label{table1}
  \end{table}

  As we proceed to higher values of odd $N$, the near perfect
  nature of the distributable entanglement within a short enough time
  (we limit our search to a range of $\beta$ from $0$ to $5000$)
  does not continue to hold. One may then ask how well one can distribute entanglement between any pair
  of parties in a multi-party scenario with the parties being
  separated from each other by at least two spins (in other words,
  each spin segment that links two parties should at least be a
  $N=4$ open spin chain with the parties at the end). The minimal ring
  for which this is possible is $N=9$. Any pair of three parties located at sites
  $1,4$ and $7$ can exchange an entanglement of $\xi\sim 0.9988$ in a scaled time $\beta=8481.4$ by
  choosing an appropriate flux. For example, to exchange between
  $1$ and $4$, we require $f=-0.25$ and between $1$ and $7$ we require
  $f=0.25$ (exchange between other combinations follow from rotational symmetry). Similarly for a ring of $N=15$, we can have a high
  entanglement distribution ($\xi \sim 0.9333$) in a scaled time of $\beta=11502$ between sites $1$ and $6$ by choosing
  $f=0.25$ and between sites $1$ and $11$ by choosing $f=-0.25$.

Let us now prove a general result of spin rings with
$N=4\mathcal{N}$ (here and following, we always let $\mathcal{N}$
to be integer). This will be the only effect of a flux which we
can prove in general (as long as $N=4\mathcal{N}$) without
choosing specific values of $N$. We examine quantum communication
between sites $1$ and $2\mathcal{N}+1$, then by choosing $f=0.5$
and Eq.~(\ref{fn}) becoming
\begin{eqnarray}
f^N_{r,s}(t) \propto \sum_{k=0}^{\infty} (-1)^k \Bigl(
J_{d+kN}(\beta) -(-1)^{d}  J_{(N-d)+kN}(\beta)\Bigr),\nonumber
\end{eqnarray}
 where $d=r-s=2\mathcal{N}$ and $N-d=d$, it's easy to find that
 $f^N_{r,s}(t)\equiv 0$ for this case.
 Thus Alice can put a flux
through the ring (which can be done locally near her end of the
ring) and control whether a Bob located at the diametrically
opposite side receives any information. We will now present an
application of this result in entangling and interfacing flux and
spin qubits without directly embedding the spin in the flux.

\begin{figure}
\begin{center}
\includegraphics[width=3in, clip]{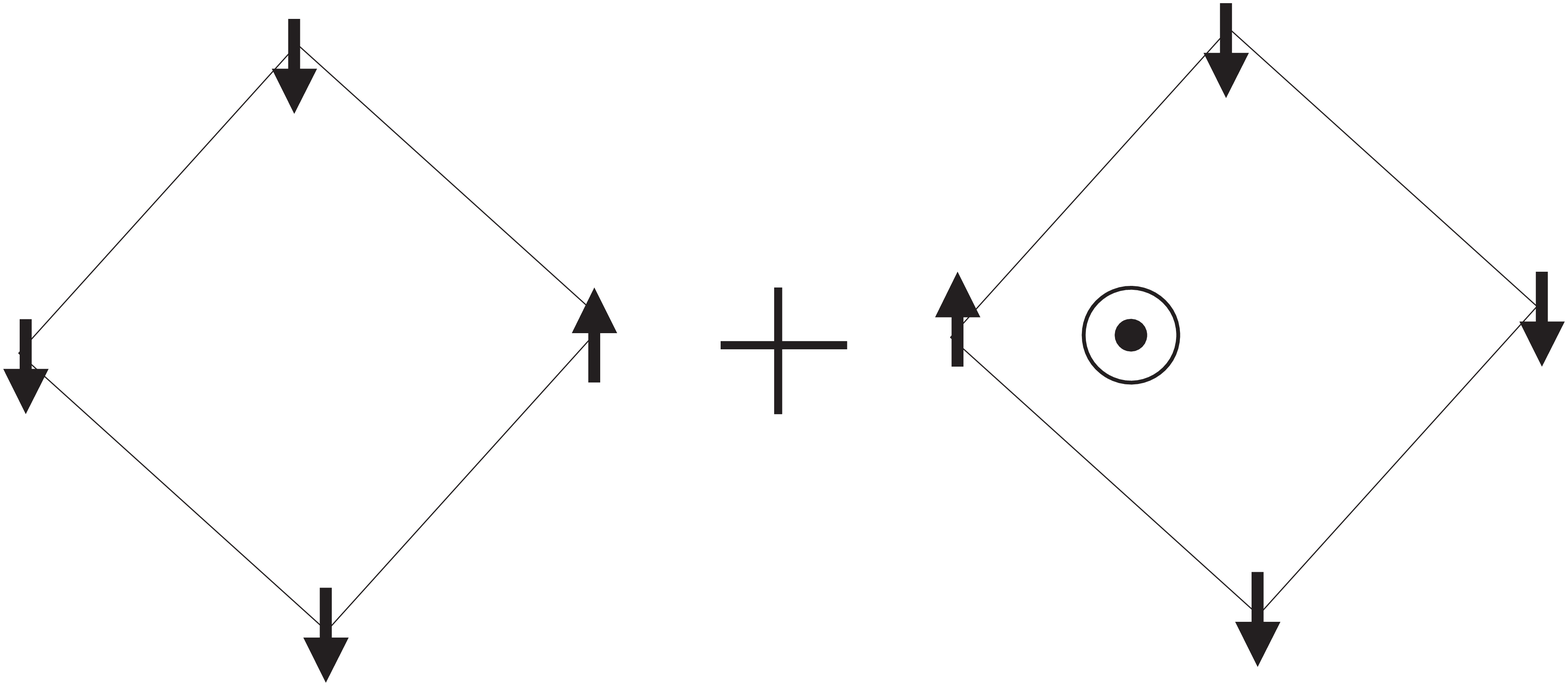}
 \caption{Entangled state generated by twisting boundary conditions. The concentric circles denote an appropriate non-zero flux and the up and down arrows denote spin states.}
\label{fluxentangle1}
\end{center}
\end{figure}

   Consider the $N=4$ ring presented in Fig.\ref{fluxentangle1}.
Consider it initially to be in the state $|1000\rangle$, where
$|1\rangle$ stands for spin up and $|0\rangle$ stands for spin
down. We know that when we have $f=0$, there is a scaled time
$\beta_1=\pi$ at which the state $|1000\rangle$ evolves to
$|0010\rangle$ \cite{Bose02}. On the other hand, if $f=0.5$, then
$|1000\rangle$ never evolves to any state with a component of
$|0010\rangle$ because of the result of our previous paragraph. In
fact it evolves back to itself in a scaled time
$\beta_2=\sqrt{2}\pi$. Therefore, if we start with the following
combined state of the flux and the spin-ring
\begin{equation}
|\Omega(0)\rangle=\frac{1}{\sqrt{2}}(|f=0\rangle +
|f=0.5\rangle)\otimes|1000\rangle,
\end{equation}
and let the state evolve for a scaled time $\beta_3=
8.5\beta_1\approx 6\beta_2$, we have the resultant spin-flux
maximally entangled state
\begin{equation}
|\Omega(\beta_3)\rangle=\frac{1}{\sqrt{2}}(|f=0\rangle
\otimes|0010\rangle+ |f=0.5\rangle \otimes|1000\rangle).
\end{equation}
This entangled state is depicted in Fig.\ref{fluxentangle1}. What
is the advantage of such a scheme for entangling spin qubits with
flux qubits as opposed to simply embedding a spin qubit in the
magnetic field of a flux qubit and letting it interact with it for
a specific time? In the case of direct embedding, one would either
have to move the spin physically or switch off the magnetic field
(of the flux) in which it is embedded, at a specific instant of
time. The advantage in our case is that the coupling between the
spins and the flux can be switched on and off merely by turning
the interactions in the spin ring on and off. There is no need to
switch off the flux or move the spins. Moreover, as the spins and
the flux do not physically reside in the same location, separate
measurements on them to verify their entanglement would be easier.
We should point out here that we do not intend to suggest an
immediately realizable experiment here, but just point out an
interesting effect which could stimulate experimental research in the direction of entangling
spin and flux qubits through boundary conditions (or in other words, the A-B effect), rather than through
interactions. Here we should point out that genuinely interesting use of the A-B effect in performing gates between 
quantum dot qubits have been proposed before \cite{pachos}, but the case presented here is very
different.

   In this paper, we have presented a scheme for multi-party quantum
 communication using a spin ring with twisted
   boundary conditions. The twist is achieved by a magnetic  flux through the
 ring. By tuning the flux appropriately, one can select any pair of parties 
along the ring to perform near perfect quantum communication, this is a new
 feature which was absent for open boundary conditions.
 We find that a 5 or 7 spin ring allows near perfect
   quantum communication between any pair of sites and higher odd N rings allow high fidelity three party quantum communication.
 An information blockage effect induced by an appropriate flux in certain even
 N rings is found. We use this result to propose a method to entangle spin and flux qubits.
 It would be interesting to investigate whether any experiment can be proposed to observe such entanglement generation between spin and flux qubits 
 through the Aharonov-Bohm effect in the future. 
 
This work was supported by  UK EPSRC Grant GR/S62796/01 and by DMR-0302758

\end{document}